\documentclass[12pt]{article}
\usepackage[a4paper, margin=1in]{geometry}  
\usepackage{graphicx}  
\usepackage{xcolor}  
\usepackage{amsmath, amssymb}  
\usepackage{longtable}  
\usepackage{booktabs}  
\usepackage{multirow}  
\usepackage{float}  
\usepackage{hyperref}  

\title{Electron Capture and $\beta$-decay Rates for Nuclei with A=65-80}
\author{
	Asim Ullah$^{1}$\thanks{Corresponding author email: \texttt{asimullah844@gmail.com}} \and
	Jameel-Un Nabi$^{1,2}$
}
\date{
	$^{1}$Faculty of Engineering Sciences, GIK Institute of Engineering Sciences and Technology, Topi 23640, Khyber Pakhtunkhwa, Pakistan \\
	$^{2}$University of Wah, Quaid Avenue, Wah Cantt 47040, Punjab, Pakistan
}

\begin{document}
	\maketitle

\begin{abstract}
{Recently a list of top 50 most important electron capture ($ec$) and $\beta$-decay ($bd$) nuclei, averaged throughout the stellar trajectory for $0.500 > {Y_e} > 0.400$, was published. The current study presents the calculation of $ec$ and $bd$ rates,  from the published list  with \textit{A} = 65--80, on a detailed temperature-density grid. The $ec$ and $bd$ rates were calculated using the proton-neutron quasiparticle random phase approximation model (pn-QRPA). Our calculation did not employ the Brink-Axel hypothesis. A systematic comparison of the current calculation with the previous pn-QRPA and independent particle model (IPM) results is presented for the first time. The reported $ec$ rates are almost the same when compared with the previous pn-QRPA calculation. On the other hand, the reported $bd$ rates are generally smaller up to an order of magnitude. Comparison with IPM results show that our calculated rates are bigger, by two orders of magnitude. The current calculation may contribute to a more realistic simulation of late phases of stellar evolution  and modeling of X-ray bursts.}

\end{abstract}
\vspace{10pt}
\noindent \textbf{Keywords}: Gamow-Teller transitions, pn-QRPA theory, Electron capture rates, $\beta$-decay rates
\section{Introduction:}
The study of supernovae is one of the ways we may learn more about our universe. In a supernova explosion, all kinds of natural interactions manifest themselves. Many scenarios of the universe are explained by the study of these interactions \cite{Sar13}. The nuclei up to Fe core are formed due to the strong and weak interactions during the hydrostatic process of stellar evolution while more massive are synthesized during the supernovae explosion at conditions with very high density and temperature. To date, the complete mechanism of supernova explosion is unknown. There are numerous complexities to consider. Researchers all over the world are working to gain a better understanding of the dynamics of core-collapse.\\
Weak interactions, especially electron capture (\textit{ec}) and beta decay (\textit{bd}), strongly influence the late evolutionary phases of heavy mass stars, determining the core entropy and electron-to-baryon ratio ($Y_e$) of the star prior to supernova and thus its Chandrasekhar mass, which is proportional to $Y_e^2$ \cite{Lan03}. These processes are also important in estimating the stellar core composition before supernova and the nucleosynthesis of neutron-rich (heavy mass) nuclei \cite{Bur57,Wal97}. Electron capture decreases the number of electrons available for pressure support, whereas beta decay does the opposite. Both these processes produce neutrinos, which escape the star at densities of $\rho$ $\leq$ 10$^{11}$ gcm$^{-3}$, carrying energy and entropy away from the core. The computation of these reactions under stellar conditions largely depend on the accurate and reliable calculation of Gamow-Teller (GT) response of the nuclei \cite{Bet79}. \\
The GT transitions are the most common spin–isospin ($\sigma$$\tau$) type weak interaction processes in atomic nuclei \cite{Ost92}. At start of core-collapse, due to low densities ($\sim$ 10$^{10}$ g/cm$^3$) and temperatures (300 - 800 keV), the chemical potential of electrons and nuclear Q-value have comparable magnitudes. Under such scenario, the \textit{ec} rates are sensitive to the detailed description of the GT strength distributions. However at much higher core densities, once the chemical potential exceeds the Q-value, the \textit{ec} rates are governed by  the total GT strength and centroid values. Therefore, a detailed knowledge of the GT distributions is in order for a reliable computation of stellar weak rates and $\beta$-decay half-lives. The GT strength distributions of hundreds of unstable nuclei are required for modeling and simulation of core-collapse supernovae. A nuclear model, preferably microscopic in nature, is required that can reliably calculate GT strength distributions and associated $\beta$-decay half-lives in reasonable agreement  with the available experimental data.   \\ 
Several attempts have been made in the past to compute weak interaction rates in stellar matter to get a better understanding of stellar dynamics. Beta decay rates in stellar matter were computed by Takahashi et al. \cite{Tak78} employing the so-called gross theory of beta decay. The drawback in this theory was that the nuclear structure details of the individual nuclei were not taken into account, instead only a statistical description of the $\beta$-strength function was assumed. The first extensive effort to tabulate the weak interaction rates at high temperatures and densities, where decays from excited states of the parent nuclei become relevant, was done by Fuller, Fowler, and Newman (FFN) \cite{Ful80,Ful82,Ful82a,Ful85}. They computed stellar electron and positron emission rates, continuum electron and positron capture rates and associated neutrino energy loss rates, for a total of 226 nuclei with masses ranging from A = 21 to 60, using the independent-particle model (IPM) with experimental data that was available at the time for astrophysical applications. Their rates led to considerable decrease in the $Y_e$ fraction throughout the core, allowing researchers to discover more about the evolution of stars prior to supernova \cite{Wea85}. In 1994, Aufderheide and colleagues \cite{Auf94} used the IPM model to examine the impact of weak interaction rates in the evolutionary phases of stars following silicon burning and provide a list of the most relevant nuclei in the $Y_e$ range of 0.40—0.50. They expanded on the FFN study by including heavy nuclei (A $>$ 60) and quenched the GT strength explicitly. Later few researchers e.g. \cite{Vet89,El-k94,Wil95} highlighted a flaw in the systematic parameterization used by FFN and Aufderheide et al. Subsequently, the proton-neutron quasiparticle random phase approximation (pn-QRPA) \cite{Nab99} and shell model \cite{Lan00} computed stellar weak rates and discovered that the GT centroids for several important nuclei were not correctly placed by FFN which resulted in a discrepancy between experimental and theoretical data. The pn-QRPA model is well suited for stellar rate computations because, unlike conventional stellar rate calculations, it does not employ the Brink-Axel hypothesis \cite{Bri58} to compute excited-state GT strength distributions. The pn-QRPA approach, using a separable interaction with a multi-$\hbar\omega$ space, makes possible a state-by-state calculation of weak interaction rates summing over Boltzmann-weighted, microscopically estimated GT strengths for all parent excited levels. This distinguishing feature of the model makes it unique amongst all calculations of stellar weak rates. The IPM was used later to perform stellar weak rate calculation for heavier nuclei with A = 65--80~\cite{Pru03}. The IPM rates has been used, in the past, in many core-collapse simulation codes. Yet there were inconsistencies found in the IPM calculation of weak rates mentioned above. Recently, Nabi and collaborators~\cite{Nab21} performed simulation study of the presupernova evolution employing the pn-QRPA computed rates. The pn-QRPA model, with optimized model parametrs and latest available measured data,  was employed to compute the stellar weak rates and mass abundances of  728 nuclei with mass in the range A = 1–-100. The authors  compiled a list of 50 most important \textit{bd} and \textit{ec} nuclei that had the greatest impact on $Y_e$.    There was a need to compare the recent pn-QRPA rates~\cite{Nab21} with the old pn-QRPA calculation~\cite{Nab99} (see also \cite{Nab04}) and the IPM rates~\cite{Pru03} for heavy nuclei (A = 65--80).

In this project, we concentrate on the top 50 most relevant $ec$ and $bd$ nuclei averaged over the entire stellar route for $Y_e$=0.40--0.50 (see Table 7 of \cite{Nab21}). Among the top 50 most relevant nuclei, we previously studied top 20 $ec$ and $bd$ \textit{fp}-shell nuclei with A $<$ 65 and compared our computations with the earlier results of large scale shell model \cite{Lan01}. In the current work, we select most important \textit{fp}-shell nuclei with A = 65--80 from the list compiled by \cite{Nab21} and compute the stellar weak rates using the pn-QRPA model.   

The paper is structured as follows. Section~2 provides a brief overview of the formalism used in our calculation. Section~3 discusses our results. The final section contains summary and concluding remarks.    
\section{Theoretical Formalism}
The weak interaction rate from the i$^{th}$ parent state to the j$^{th}$ daughter state in stellar matter can be computed via the formula given below
\begin{eqnarray}\label{rate}
	\lambda^{ec(bd)}_{ij} =\frac{ln2}{D}[{\phi_{ij}^{ec(bd)}(\rho, T, E_{f})}] \times \left[B(F)_{ij}+\frac{B(GT)_{ij}}{(g_{A}/g_{V})^{-2}} \right].
\end{eqnarray}
The value of constant D appearing in Eq.~(\ref{rate}) was taken as 6143 $s$ \cite{Har09}. The value of $g_{A}$/$g_{v}$, representing the ratio of axial and vector coupling constant, was taken as -1.2694 \cite{Nak10}. $B(F)_{ij}$ and $B(GT)_{ij}$ are the total reduced transition probabilities due to Fermi and GT interactions, respectively. 

\begin{equation}
	B(F)_{ij} = [{2J_{i}+1}]^{-1}  \mid<j \parallel \sum_{k}t_{\pm}^{k}
	\parallel i> \mid ^{2}
\end{equation}

\begin{equation}\label{bgt}
	B(GT)_{ij} = [{2J_{i}+1}]^{-1}  \mid <j
	\parallel \sum_{k}t_{\pm}^{k}\vec{\sigma}^{k} \parallel i> \mid ^{2},
\end{equation}
where $J_i$, $t_\pm$ and $\vec{\sigma}$ stands for the total spin of parent nucleus, isospin operators (raising/lowering) and Pauli spin matrices, respectively. The GT strength parameters were tuned so that the computations reproduced the measured $\beta$--decay half-lives~\cite{Aud21}. For the choice of pn-QRPA model parameters, used in the current calculation, we refer to~\cite{Far21}.

The phase space integrals $(\phi_{ij})$ for \textit{ec} and \textit{bd} are; (hereafter natural units are used, $\hbar=c=m_{e}=1$)
\begin{equation}\label{pc}
	\phi^{ec}_{ij} = \int_{w_{l}}^{\infty} w (w_{m}+w)^{2}({w^{2}-1})^{\frac{1}{2}} F(+Z,w) G_{-} dw,
\end{equation}

\begin{equation}\label{ps}
	\phi^{bd}_{ij} = \int_{1}^{w_{m}} w (w_{m}-w)^{2}({w^{2}-1})^{\frac{1}{2}} F(+Z,w)
	(1-G_{-}) dw,
\end{equation}
In Eqs.~(\ref{pc}) and (\ref{ps}) $w$, $w_l$ and $w_{m}$ stands for total energy (kinetic+rest mass) of the electron, total threshold energy for \textit{ec} and the total energy of \textit{bd}, respectively. The Fermi functions ($F(+ Z, w)$) were calculated employing the procedure used by Gove and Martin ~\cite{Gov71}. $G_{-}$ represents the electron distribution function.
The \textit{ec} and \textit{bd} total weak-interaction rates were calculated using
\begin{equation}\label{ecbd}
	\lambda^{ec(bd)} =\sum _{ij}P_{i} \lambda^{ec(bd)} _{ij}.
\end{equation}
The summation was applied over all the initial and final states and satisfactory convergence in \textit{ec} and \textit{bd} rates was achieved. $P_{i} $ in  Eq.~(\ref{ecbd}) denotes the occupation probability of parent excited states and follows the normal Boltzmann distribution. 
\section{Results and Discussion}
In this work, we present the microscopic calculation of stellar weak rates for a total of 24 astrophysically most relevant \textit{ec} and \textit{bd}  nuclei with A = 65--80 using the pn-QRPA theory. These nuclei were selected from the compiled list of top 50 most relevant nuclei (see Table~7 of \cite{Nab21}) and are reproduced in Table~\ref{T1}. The \textit{ec} and \textit{bd} nuclei are shown separately. The table also displays the rank of these nuclei, adopted from \cite{Nab21}, and computed as 
\begin{equation}
	\mathring{R}_{p} = \left(\frac{\dot{Y}^{ec(bd)}_{e(i)}}{\sum\dot{Y}^{ec(bd)}_{e(i)}}\right)_{0.500 > {Y_e} > 0.400}	
\end{equation}
Thus, the nuclei that contribute the most to $\dot{Y_e}$ will have the highest $\mathring{R}_p$ value. The stellar rates (\textit{ec} and \textit{bd}) were calculated microscopically using the pn-QRPA model incorporating the most recent experimental data and optimal values of model parameter. We made a comparison of our computed rates (hereafter this calculation will also be referred to as pnQRPA-21) with the previous pn-QRPA calculation~\cite{Nab04} (hereafter referred to as pn-QRPA-04) and IPM rates~\cite{Pru03} (hereafter referred to as IPM-03).
Tables~(\ref{T1ec}-\ref{T4ec}) and Tables~(\ref{T1bd}-\ref{T4bd}) compare our calculated \textit{ec} and \textit{bd} rates, respectively, with those of pn-QRPA-04 and IPM-03 results. It is to be noted that exponents are shown in parenthesis in all tables.  Tables~(\ref{T1ec}~-~\ref{T4ec}) show the comparison of computed \textit{ec} rates for  13 most relevant \textit{ec} nuclei at core temperatures [$T$ = (1, 2, 3, 5, 10 \& 30) GK] and densities [$\rho$$Y_e$ = ($10^{1}-10^{11}$) g/cm$^3$].  Tables~(\ref{T1bd}~-~\ref{T4bd}) show a similar comparison for \textit{bd} rates. These tables demonstrate that, as the stellar temperature increases, so do the \textit{ec} and \textit{bd} rates on the chosen nuclei. This is because the occupation probability of the parent excited states rises as the temperature is raised, contributing efficiently to the total rates. The \textit{ec} rates rise as $\rho$$Y_e$ values increase due to a corresponding rise in electron chemical potential. The \textit{bd} rates, on the other hand, are reduced for high $\rho$$Y_e$ values due to a decrease in available phase space caused by increased electron chemical potential. \\

A comparison of the pn-QRPA-21 and pn-QRPA-04 calculations, shown in Tables~(\ref{T1ec}--\ref{T4ec}), reveal that the $ec$ rates for these neutron-rich nuclei are negligible at small temperature and density values and would hardly have any affect on the simulation results. Even at high temperatures and densities, the two calculations are in decent agreement with few minor changes.  There are two main reasons for these minor differences. The first is the choice of different GT strength parameters in the two pn-QRPA calculations. Instead of using fixed values of $\kappa$ and $\chi$ (GT interaction constants) for each isotopic chain (as done in \cite{Nab04}), the constants values were calculated using 0.25/A$^{0.7}$ and 2.8/A$^{0.7}$, respectively, adopted from Ref. \cite{Hom96}. Yet another source for change in the old and new pn-QRPA calculations is the Q-values used for the calculation of weak rates. Whereas the pn-QRPA-04 used Q-values from Ref~\cite{Aud95}, the recent pn-QRPA-21 employed Q-values from Ref.~\cite{Aud17}.

The comparison of $bd$ rates between pn-QRPA-21 and pn-QRPA-04 calculations (see Tables~(\ref{T1bd}--\ref{T4bd})) suggests that  the two calculations are comparable for most cases. However, the pn-QRPA21 $bd$ rates are generally suppressed, by up to an order of magnitude at high temperatures for all densities. The differences are attributed to the two reasons mentioned above.    

The interesting question is how the pnQRPA-21 calculation compares with the IPM-03 results used in many simulation codes. The pn-QRPA-21 calculated \textit{ec} and $bd$ rates, for almost all the cases shown in Tables~(\ref{T1ec}--\ref{T4bd}), are bigger than the corresponding IPM-03 rates. The pn-QRPA-21 $ec$ rates are bigger by up to two orders of magnitude at high temperatures. At low core temperatures and densities, the $ec$ rates are too small for the considered neutron-rich nuclei to have any impact on the simulations results. The pn-QRPA-04 \textit{bd} rates, shown in Tables~(\ref{T1bd}--\ref{T4bd}),  are, in general, found to be bigger (up to two orders of magnitude) than IPM-03 rates at all $T_9$ and $\rho$$Y_e$ values.  
The PF03 incorporated the Brink's hypothesis \cite{Bri58} to measure contribution from higher excited states at high density and temperature values, which is not a good approximation to be used in such calculation~\cite{Nab21+}. The PF03 approach used a quenching factor of 4 and 3 for GT$^+$ and GT$^-$ transitions, respectively. We did not incorporate any quenching factor in our calculation. The PF03 calculation included many states in the partition function sum without including the corresponding weak interaction strength.  These may be cited as key reasons for the smaller PF03 rates especially at high $\rho$$Y_e$ and $T_9$ values.  The pn-QRPA approach, on the other hand, made use of a large model space (up to 7$\hbar \omega$) to adequately handle excited states in heavy nuclei (courtesy of the separable schematic interaction). We again note that Brink's hypothesis was not used in computation of excited state GT strength functions in our calculation \cite{Nab21+}. 

\section{Summary and Conclusion} 
In this work we present the microscopic computation of stellar weak rates for 24 neutron-rich heavy nuclei of astrophysical significance. We calculated  \textit{ec} and \textit{bd} rates for nuclei with \textit{A} = 65--80 using the pn-QRPA model. The selected nuclei contribute significantly to change in $Y_e$ values according to a recent survey~\cite{Nab21}. The stellar weak rates (\textit{ec} and \textit{bd}) were computed by summing over microscopically estimated GT strength for all parent excited states. These rates were computed for a broad range of temperatures (0.1 - 30) GK and densities ($10-10^{11}$) g/cm$^3$ and were later compared with the original pn-QRPA calculation~\cite{Nab04} as well as the frequently employed IPM rates~\cite{Pru03}. The old and new pn-QRPA $ec$ rates are in decent comparison whereas the new $bd$ rates are generally smaller by up to an order of magnitude at high temperatures. The pn-QRPA-21 computed \textit{ec} and \textit{bd} rates are generally bigger than IPM-03 rates by up to two orders of magnitude and may bear consequences. Quenching factors, inaccurate determination of nuclear partition functions and usage of Brink's hypothesis could be the probable reasons of smaller IPM-F03 rates, especially at high $T_9$ and $\rho$$Y_e$ values.  The current study may prove useful for simulation of later phases of stellar evolution processes and modeling of X-ray bursts. A detailed calculation for proton-rich nuclei with A = 80 - 110 is currently in progress and we hope to report our findings in near future.

%

\newpage

\newpage
\begin{table*}[]
	\caption{Most important \textit{fp}-shell nuclei with A = 65--80 extracted from Ref.~\cite{Nab21} and considered for calculation in this project. The electron capture (\textit{ec}) and $\beta$-decay (\textit{bd}) nuclei are shown in separate columns.}
	\label{T1}
	\centering
	\begin{tabular}{ccc|ccc}
		
		\hline
		\multicolumn{3}{c}{\textit{ec} nuclei} & \multicolumn{3}{c}{\textit{bd} nuclei}   \\
		
		\hline
		A   & Element  & $\mathring{R}_{p}$  & 	A   & Element  & $\mathring{R}_{p}$   \\
		67 & Cu & 3  & 67 & Ni & 1  \\
		66 & Cu & 7  & 69 & Cu & 12 \\
		79 & Ge & 11 & 67 & Co & 25 \\
		77 & Ga & 12 & 66 & Co & 27 \\
		78 & Ge & 14 & 71 & Cu & 29 \\
		67 & Ni & 17 & 77 & Ga & 32 \\
		68 & Cu & 19 & 75 & Ga & 35 \\
		75 & Ga & 24 & 79 & Ga & 36 \\
		69 & Cu & 26 & 70 & Cu & 37 \\
		78 & Ga & 29 & 68 & Ni & 47 \\
		77 & Ge & 35 & 68 & Co & 49 \\
		71 & Cu & 46 &    &    &    \\
		73 & Ga & 49 &    &    &   \\
		\hline
	\end{tabular}
\end{table*}

\begin{table*}[] 
	\caption{The \textit{ec} rates ($\lambda^{ec}$) on $^{67}$Cu, $^{66}$Cu, $^{79}$Ge, $^{77}$Ga, $^{78}$Ge, $^{67}$Ni and $^{68}$Cu. The previous pn-QRPA rates~\cite{Nab04} and IPM calculation~\cite{Pru03} are also shown for comparison. The units of $T_9$, $\rho$$\it Y_{e}$ and $\lambda^{ec}$ are GK, g/cm$^3$ and $s^{-1}$, respectively. The nuclei are ordered on the basis of ranking shown in Table~\ref{T1}. Number in parenthesis denotes exponents. In the table, $1.00(-100)$ means that the rate is smaller than 10$^{-100} (s^{-1})$. }
	\label{T1ec}
	\centering
	 \resizebox{\textwidth}{!}{
		\begin{tabular}{c|c|ccc|ccc|ccc|}
			\hline
			\multicolumn{5}{l}{} & \multicolumn{1}{l}{$\lambda^{ec}$} & \multicolumn{5}{l}{}        \\
			\cline{1-11} \multicolumn{1}{l}{} &  & \multicolumn{3}{c}{$\rho$$\it Y_{e}$ = $10^1$ }        & \multicolumn{3}{c}{$\rho$$\it Y_{e}$ = $10^3$ }& \multicolumn{3}{c}{$\rho$$\it Y_{e}$ = $10^5$ }         \\
			\cline{3-11} \multicolumn{1}{l}{Nuclei} &       $T_9$                     & pn-QRPA-21& pn-QRPA-04    & IPM-03      & pn-QRPA-21 &pn-QRPA-04  & IPM-03      & pn-QRPA-21&pn-QRPA-04   & IPM-03 \\
			\hline
			\\{$^{67}$Cu} & 1  & 3.34(-25) & 2.27(-25) & 1.01(-25) & 5.65(-25) & 3.84(-25) & 1.70(-25) & 4.03(-23) & 2.73(-23) & 1.21(-23) \\
			& 2  & 7.46(-14) & 5.45(-14) & 1.91(-14) & 7.52(-14) & 5.50(-14) & 1.93(-14) & 1.57(-13) & 1.15(-13) & 4.03(-14) \\
			& 3  & 8.49(-10) & 6.53(-10) & 2.24(-10) & 8.49(-10) & 6.55(-10) & 2.24(-10) & 9.73(-10) & 7.50(-10) & 2.57(-10) \\
			& 5  & 2.93(-06) & 2.58(-06) & 9.42(-07) & 2.93(-06) & 2.58(-06) & 9.42(-07) & 2.99(-06) & 2.64(-06) & 9.62(-07) \\
			& 10 & 7.41(-03) & 1.58(-02) & 2.25(-03) & 7.41(-03) & 1.58(-02) & 2.25(-03) & 7.43(-03) & 1.58(-02) & 2.25(-03) \\
			& 30 & 5.41(+02) & 4.66(+02) & 2.43(+01) & 5.42(+02) & 4.66(+02) & 2.43(+01) & 5.42(+02) & 4.66(+02) & 2.43(+01) \\
			\\{$^{66}$Cu} & 1  & 7.41(-24) & 7.14(-10) & 2.17(-10) & 1.25(-23) & 1.21(-09) & 3.66(-10) & 8.93(-22) & 8.61(-08) & 2.60(-08) \\
			& 2  & 2.92(-13) & 4.56(-07) & 1.10(-07) & 2.94(-13) & 4.60(-07) & 1.11(-07) & 6.14(-13) & 9.59(-07) & 2.31(-07) \\
			& 3  & 2.01(-09) & 7.14(-06) & 1.54(-06) & 2.02(-09) & 7.16(-06) & 1.54(-06) & 2.31(-09) & 8.20(-06) & 1.76(-06) \\
			& 5  & 6.31(-06) & 2.46(-04) & 3.04(-05) & 6.31(-06) & 2.46(-04) & 3.05(-05) & 6.44(-06) & 2.51(-04) & 3.11(-05) \\
			& 10 & 1.78(-02) & 2.21(-01) & 3.58(-03) & 1.78(-02) & 2.21(-01) & 3.58(-03) & 1.78(-02) & 2.21(-01) & 3.59(-03) \\
			& 30 & 5.71(+02) & 8.13(+02) & 5.57(+01) & 5.71(+02) & 8.13(+02) & 5.57(+01) & 5.71(+02) & 8.15(+02) & 5.57(+01) \\
			\\{$^{79}$Ge} & 1  & 1.19(-43) & 9.44(-43) & 7.41(-43) & 2.01(-43) & 1.60(-42) & 1.25(-42) & 1.44(-41) & 1.14(-40) & 8.89(-41) \\
			& 2  & 7.89(-24) & 6.08(-23) & 5.13(-23) & 7.96(-24) & 6.12(-23) & 5.16(-23) & 1.66(-23) & 1.28(-22) & 1.08(-22) \\
			& 3  & 1.74(-16) & 5.55(-16) & 4.16(-16) & 1.75(-16) & 5.55(-16) & 4.17(-16) & 2.00(-16) & 6.35(-16) & 4.76(-16) \\
			& 5  & 1.03(-09) & 1.41(-09) & 3.92(-10) & 1.03(-09) & 1.41(-09) & 3.92(-10) & 1.05(-09) & 1.44(-09) & 4.00(-10) \\
			& 10 & 5.98(-04) & 8.11(-04) & 6.17(-05) & 5.98(-04) & 8.11(-04) & 6.17(-05) & 6.00(-04) & 8.13(-04) & 6.18(-05) \\
			& 30 & 3.45(+02) & 4.28(+02) & 9.51(+00) & 3.45(+02) & 4.28(+02) & 9.51(+00) & 3.45(+02) & 4.28(+02) & 9.51(+00) \\
			\\{$^{77}$Ga} & 1  & 2.26(-43) & 1.07(-43) & 9.59(-45) & 3.83(-43) & 1.81(-43) & 1.62(-44) & 2.74(-41) & 1.29(-41) & 1.15(-42) \\
			& 2  & 6.73(-23) & 4.79(-23) & 5.51(-24) & 6.79(-23) & 4.83(-23) & 5.55(-24) & 1.42(-22) & 1.01(-22) & 1.16(-23) \\
			& 3  & 1.13(-15) & 9.62(-16) & 1.06(-16) & 1.13(-15) & 9.62(-16) & 1.06(-16) & 1.29(-15) & 1.10(-15) & 1.21(-16) \\
			& 5  & 2.58(-09) & 2.48(-09) & 1.88(-10) & 2.58(-09) & 2.48(-09) & 1.88(-10) & 2.64(-09) & 2.53(-09) & 1.92(-10) \\
			& 10 & 1.00(-03) & 9.93(-04) & 4.41(-05) & 1.00(-03) & 9.93(-04) & 4.41(-05) & 1.00(-03) & 9.95(-04) & 4.42(-05) \\
			& 30 & 5.08(+02) & 4.67(+02) & 8.93(+00) & 5.08(+02) & 4.68(+02) & 8.93(+00) & 5.08(+02) & 4.68(+02) & 8.93(+00) \\
			\\{$^{78}$Ge} & 1  & 5.20(-49) & 1.29(-48) & 1.80(-48) & 8.79(-49) & 2.18(-48) & 3.03(-48) & 6.28(-47) & 1.56(-46) & 2.15(-46) \\
			& 2  & 4.65(-25) & 5.42(-25) & 2.90(-25) & 4.69(-25) & 5.46(-25) & 2.93(-25) & 9.79(-25) & 1.14(-24) & 6.11(-25) \\
			& 3  & 9.33(-17) & 9.48(-17) & 1.82(-17) & 9.35(-17) & 9.51(-17) & 1.82(-17) & 1.07(-16) & 1.09(-16) & 2.09(-17) \\
			& 5  & 7.73(-10) & 6.84(-10) & 5.42(-11) & 7.74(-10) & 6.84(-10) & 5.42(-11) & 7.91(-10) & 6.98(-10) & 5.53(-11) \\
			& 10 & 4.99(-04) & 3.48(-04) & 2.08(-05) & 5.00(-04) & 3.48(-04) & 2.08(-05) & 5.01(-04) & 3.48(-04) & 2.09(-05) \\
			& 30 & 1.73(+02) & 5.38(+01) & 5.87(+00) & 1.73(+02) & 5.40(+01) & 5.87(+00) & 1.73(+02) & 5.40(+01) & 5.87(+00) \\
			\\{$^{67}$Ni} & 1  & 1.10(-48) & 8.09(-49) & 2.82(-49) & 1.85(-48) & 1.37(-48) & 4.75(-49) & 1.32(-46) & 9.77(-47) & 3.38(-47) \\
			& 2  & 4.91(-25) & 4.53(-25) & 8.53(-26) & 4.94(-25) & 4.57(-25) & 8.59(-26) & 1.03(-24) & 9.55(-25) & 1.79(-25) \\
			& 3  & 9.55(-17) & 9.44(-17) & 8.81(-18) & 9.55(-17) & 9.46(-17) & 8.83(-18) & 1.09(-16) & 1.08(-16) & 1.01(-17) \\
			& 5  & 8.87(-10) & 8.91(-10) & 3.82(-11) & 8.89(-10) & 8.91(-10) & 3.82(-11) & 9.08(-10) & 9.10(-10) & 3.90(-11) \\
			& 10 & 6.41(-04) & 6.38(-04) & 1.77(-05) & 6.41(-04) & 6.40(-04) & 1.77(-05) & 6.43(-04) & 6.41(-04) & 1.78(-05) \\
			& 30 & 3.44(+02) & 3.30(+02) & 5.40(+00) & 3.44(+02) & 3.30(+02) & 5.40(+00) & 3.44(+02) & 3.30(+02) & 5.40(+00) \\
			\\{$^{68}$Cu} & 1  & 1.12(-31) & 4.41(-30) & 2.25(-17) & 1.90(-31) & 7.45(-30) & 3.79(-17) & 1.36(-29) & 5.32(-28) & 2.70(-15) \\
			& 2  & 8.07(-17) & 1.63(-15) & 4.71(-10) & 8.15(-17) & 1.65(-15) & 4.74(-10) & 1.70(-16) & 3.44(-15) & 9.91(-10) \\
			& 3  & 1.32(-11) & 2.10(-10) & 2.15(-07) & 1.33(-11) & 2.10(-10) & 2.16(-07) & 1.52(-11) & 2.41(-10) & 2.47(-07) \\
			& 5  & 4.18(-07) & 5.06(-06) & 5.32(-05) & 4.18(-07) & 5.06(-06) & 5.32(-05) & 4.27(-07) & 5.16(-06) & 5.43(-05) \\
			& 10 & 6.32(-03) & 3.72(-02) & 1.38(-02) & 6.32(-03) & 3.72(-02) & 1.38(-02) & 6.34(-03) & 3.72(-02) & 1.39(-02) \\
			& 30 & 5.22(+02) & 5.97(+02) & 4.82(+01) & 5.24(+02) & 5.98(+02) & 4.82(+01) & 5.24(+02) & 5.98(+02) & 4.82(+01) \\
			\\ \hline
	\end{tabular}}
\end{table*}
\begin{table*}[]
	\caption{Same as Table~\ref{T1ec} but for higher values of  density.}
	\label{T2ec}
	\centering
		\centering
	\resizebox{\textwidth}{!}{
		\begin{tabular}{c|c|ccc|ccc|ccc|}
			\hline
			\multicolumn{5}{l}{} & \multicolumn{1}{l}{$\lambda^{ec}$} & \multicolumn{5}{l}{}        \\
			\cline{1-11} \multicolumn{1}{l}{} &  & \multicolumn{3}{c}{$\rho$$\it Y_{e}$ = $10^7$ }        & \multicolumn{3}{c}{$\rho$$\it Y_{e}$ = $10^9$ }& \multicolumn{3}{c}{$\rho$$\it Y_{e}$ = $10^{11}$ }         \\
			\cline{3-11} \multicolumn{1}{l}{Nuclei} &       $T_9$                     & pn-QRPA-21& pn-QRPA-04    & IPM-03      & pn-QRPA-21 &pn-QRPA-04  & IPM-03      & pn-QRPA-21&pn-QRPA-04   & IPM-03 \\
			\hline
			\\{$^{67}$Cu} & 1  & 3.72(-19) & 2.37(-19) & 1.11(-19) & 2.51(-02) & 2.04(-02) & 1.49(-02) & 3.23(+04) & 2.86(+04) & 8.85(+03) \\
			& 2  & 5.32(-11) & 3.61(-11) & 1.36(-11) & 3.92(-02) & 3.03(-02) & 1.94(-02) & 3.70(+04) & 3.24(+04) & 8.87(+03) \\
			& 3  & 4.39(-08) & 3.23(-08) & 1.16(-08) & 5.74(-02) & 4.36(-02) & 2.80(-02) & 3.85(+04) & 3.40(+04) & 8.75(+03) \\
			& 5  & 1.48(-05) & 1.28(-05) & 4.75(-06) & 1.28(-01) & 9.59(-02) & 6.70(-02) & 3.93(+04) & 3.52(+04) & 8.75(+03) \\
			& 10 & 9.31(-03) & 1.99(-02) & 2.82(-03) & 1.30(+00) & 2.88(+00) & 4.58(-01) & 5.64(+04) & 6.49(+04) & 8.89(+03) \\
			& 30 & 5.46(+02) & 4.70(+02) & 2.44(+01) & 1.15(+03) & 9.89(+02) & 5.02(+01) & 7.28(+05) & 5.38(+05) & 1.74(+04) \\
			\\{$^{66}$Cu} & 1  & 8.26(-18) & 1.84(-04) & 5.53(-05) & 2.75(-02) & 2.52(+00) & 5.02(-01) & 5.02(+04) & 1.34(+05) & 1.02(+04) \\
			& 2  & 2.07(-10) & 2.33(-04) & 5.58(-05) & 6.22(-02) & 3.30(+00) & 3.13(-01) & 7.41(+04) & 2.74(+05) & 9.53(+03) \\
			& 3  & 1.03(-07) & 3.26(-04) & 7.01(-05) & 1.11(-01) & 4.62(+00) & 2.49(-01) & 8.63(+04) & 3.46(+05) & 9.14(+03) \\
			& 5  & 3.16(-05) & 1.22(-03) & 1.50(-04) & 3.08(-01) & 9.64(+00) & 2.48(-01) & 9.77(+04) & 4.11(+05) & 8.99(+03) \\
			& 10 & 2.23(-02) & 2.77(-01) & 4.49(-03) & 3.13(+00) & 4.55(+01) & 6.35(-01) & 1.25(+05) & 4.81(+05) & 9.08(+03) \\
			& 30 & 5.77(+02) & 8.20(+02) & 5.61(+01) & 1.22(+03) & 1.73(+03) & 1.14(+02) & 7.94(+05) & 9.89(+05) & 2.96(+04) \\
			\\{$^{79}$Ge} & 1  & 1.32(-37) & 1.05(-36) & 8.18(-37) & 1.26(-17) & 4.98(-18) & 8.41(-17) & 3.44(+04) & 3.44(+04) & 4.81(+03) \\
			& 2  & 5.61(-21) & 4.36(-20) & 3.66(-20) & 7.08(-11) & 1.59(-10) & 4.41(-10) & 3.48(+04) & 3.53(+04) & 4.83(+03) \\
			& 3  & 9.04(-15) & 2.88(-14) & 2.16(-14) & 7.14(-08) & 1.38(-07) & 1.65(-07) & 3.61(+04) & 3.81(+04) & 4.79(+03) \\
			& 5  & 5.20(-09) & 7.11(-09) & 1.99(-09) & 1.29(-04) & 1.67(-04) & 4.85(-05) & 3.89(+04) & 4.48(+04) & 4.83(+03) \\
			& 10 & 7.52(-04) & 1.02(-03) & 7.74(-05) & 1.41(-01) & 1.90(-01) & 1.44(-02) & 5.55(+04) & 7.16(+04) & 5.01(+03) \\
			& 30 & 3.48(+02) & 4.32(+02) & 9.57(+00) & 7.38(+02) & 9.14(+02) & 1.97(+01) & 6.50(+05) & 7.98(+05) & 9.25(+03) \\
			\\{$^{77}$Ga} & 1  & 2.53(-37) & 1.20(-37) & 1.06(-38) & 1.98(-18) & 1.20(-18) & 1.06(-18) & 3.71(+04) & 3.71(+04) & 4.41(+03) \\
			& 2  & 4.83(-20) & 3.44(-20) & 3.94(-21) & 2.10(-10) & 1.67(-10) & 5.27(-11) & 3.81(+04) & 3.80(+04) & 4.47(+03) \\
			& 3  & 5.86(-14) & 4.99(-14) & 5.50(-15) & 2.88(-07) & 2.61(-07) & 4.41(-08) & 4.04(+04) & 4.03(+04) & 4.41(+03) \\
			& 5  & 1.30(-08) & 1.25(-08) & 9.51(-10) & 2.90(-04) & 2.83(-04) & 2.36(-05) & 4.47(+04) & 4.46(+04) & 4.45(+03) \\
			& 10 & 1.26(-03) & 1.25(-03) & 5.53(-05) & 2.29(-01) & 2.30(-01) & 1.03(-02) & 6.84(+04) & 6.95(+04) & 4.61(+03) \\
			& 30 & 5.12(+02) & 4.71(+02) & 9.02(+00) & 1.09(+03) & 1.00(+03) & 1.85(+01) & 9.02(+05) & 8.53(+05) & 8.73(+03) \\
			\\{$^{78}$Ge} & 1  & 5.79(-43) & 1.44(-42) & 1.98(-42) & 3.44(-23) & 3.46(-23) & 2.10(-22) & 3.18(+04) & 3.19(+04) & 3.78(+03) \\
			& 2  & 3.33(-22) & 3.89(-22) & 2.08(-22) & 4.51(-12) & 4.40(-12) & 2.30(-12) & 3.18(+04) & 3.11(+04) & 3.79(+03) \\
			& 3  & 4.85(-15) & 4.93(-15) & 9.46(-16) & 3.91(-08) & 3.61(-08) & 6.67(-09) & 3.19(+04) & 2.93(+04) & 3.75(+03) \\
			& 5  & 3.91(-09) & 3.45(-09) & 2.74(-10) & 9.68(-05) & 8.04(-05) & 6.44(-06) & 3.23(+04) & 2.61(+04) & 3.78(+03) \\
			& 10 & 6.27(-04) & 4.37(-04) & 2.62(-05) & 1.17(-01) & 7.98(-02) & 4.69(-03) & 4.70(+04) & 3.02(+04) & 3.94(+03) \\
			& 30 & 1.75(+02) & 5.43(+01) & 5.92(+00) & 3.71(+02) & 1.15(+02) & 1.22(+01) & 3.27(+05) & 1.01(+05) & 7.00(+03) \\
			\\{$^{67}$Ni} & 1  & 1.22(-42) & 9.04(-43) & 3.11(-43) & 1.30(-22) & 8.36(-23) & 3.30(-23) & 4.29(+03) & 1.98(+03) & 3.24(+03) \\
			& 2  & 3.52(-22) & 3.26(-22) & 6.10(-23) & 4.93(-12) & 4.31(-12) & 8.53(-13) & 4.92(+03) & 3.47(+03) & 3.32(+03) \\
			& 3  & 4.97(-15) & 4.92(-15) & 4.58(-16) & 4.06(-08) & 3.92(-08) & 3.74(-09) & 6.10(+03) & 6.98(+03) & 3.33(+03) \\
			& 5  & 4.49(-09) & 4.51(-09) & 1.93(-10) & 1.12(-04) & 1.11(-04) & 4.81(-06) & 1.11(+04) & 1.52(+04) & 3.43(+03) \\
			& 10 & 8.05(-04) & 8.02(-04) & 2.22(-05) & 1.51(-01) & 1.50(-01) & 3.91(-03) & 3.45(+04) & 3.75(+04) & 3.59(+03) \\
			& 30 & 3.47(+02) & 3.33(+02) & 5.45(+00) & 7.36(+02) & 7.06(+02) & 1.12(+01) & 5.98(+05) & 5.61(+05) & 6.49(+03) \\
			\\{$^{68}$Cu} & 1  & 1.25(-25) & 4.92(-24) & 2.47(-11) & 4.65(-06) & 3.33(-04) & 2.33(+00) & 4.29(+04) & 9.93(+04) & 2.98(+04) \\
			& 2  & 5.79(-14) & 1.17(-12) & 3.37(-07) & 3.18(-04) & 1.04(-02) & 1.87(+00) & 5.31(+04) & 1.05(+05) & 2.48(+04) \\
			& 3  & 6.89(-10) & 1.09(-08) & 1.12(-05) & 2.32(-03) & 6.52(-02) & 1.63(+00) & 6.34(+04) & 1.26(+05) & 2.15(+04) \\
			& 5  & 2.11(-06) & 2.56(-05) & 2.69(-04) & 2.76(-02) & 5.40(-01) & 1.51(+00) & 8.04(+04) & 1.77(+05) & 1.81(+04) \\
			& 10 & 7.94(-03) & 4.67(-02) & 1.74(-02) & 1.28(+00) & 8.20(+00) & 2.49(+00) & 1.19(+05) & 2.61(+05) & 1.56(+04) \\
			& 30 & 5.27(+02) & 6.03(+02) & 4.85(+01) & 1.12(+03) & 1.27(+03) & 9.95(+01) & 8.13(+05) & 8.87(+05) & 2.98(+04) \\
			\\\hline
	\end{tabular}}
\end{table*}

\begin{table*}[]
	\caption{Same as Table~\ref{T1ec} but for $^{75}$Ga, $^{69}$Cu, $^{78}$Ga, $^{77}$Ge, $^{71}$Cu and $^{73}$Ga.}
	\label{T3ec}
	\centering
	\centering
\resizebox{\textwidth}{!}{ 
		\begin{tabular}{c|c|ccc|ccc|ccc|}
			\hline
			\multicolumn{5}{l}{} & \multicolumn{1}{l}{$\lambda^{ec}$} & \multicolumn{5}{l}{}        \\
			\cline{1-11} \multicolumn{1}{l}{} &  & \multicolumn{3}{c}{$\rho$$\it Y_{e}$ = $10^1$ }        & \multicolumn{3}{c}{$\rho$$\it Y_{e}$ = $10^3$ }& \multicolumn{3}{c}{$\rho$$\it Y_{e}$ = $10^5$ }         \\
			\cline{3-11} \multicolumn{1}{l}{Nuclei} &       $T_9$                     & pn-QRPA-21& pn-QRPA-04    & IPM-03      & pn-QRPA-21 &pn-QRPA-04  & IPM-03      & pn-QRPA-21&pn-QRPA-04   & IPM-03 \\
			\hline
			\\{$^{75}$Ga} & 1  & 5.22(-37) & 1.96(-37) & 1.19(-38) & 8.83(-37) & 3.31(-37) & 2.01(-38) & 6.30(-35) & 2.37(-35) & 1.44(-36) \\
			& 2  & 1.07(-19) & 6.47(-20) & 5.74(-21) & 1.08(-19) & 6.53(-20) & 5.79(-21) & 2.25(-19) & 1.36(-19) & 1.21(-20) \\
			& 3  & 1.27(-13) & 8.59(-14) & 1.00(-14) & 1.27(-13) & 8.59(-14) & 1.00(-14) & 1.45(-13) & 9.84(-14) & 1.15(-14) \\
			& 5  & 1.89(-08) & 1.36(-08) & 2.63(-09) & 1.89(-08) & 1.36(-08) & 2.63(-09) & 1.93(-08) & 1.39(-08) & 2.69(-09) \\
			& 10 & 1.49(-03) & 1.04(-03) & 1.51(-04) & 1.49(-03) & 1.04(-03) & 1.51(-04) & 1.50(-03) & 1.04(-03) & 1.51(-04) \\
			& 30 & 5.09(+02) & 1.72(+02) & 1.22(+01) & 5.09(+02) & 1.72(+02) & 1.22(+01) & 5.09(+02) & 1.72(+02) & 1.22(+01) \\
			\\{$^{69}$Cu} & 1  & 1.35(-35) & 1.12(-33) & 5.69(-34) & 2.28(-35) & 1.89(-33) & 9.59(-34) & 1.62(-33) & 1.35(-31) & 6.82(-32) \\
			& 2  & 7.19(-19) & 9.04(-18) & 3.46(-18) & 7.24(-19) & 9.10(-18) & 3.48(-18) & 1.52(-18) & 1.91(-17) & 7.26(-18) \\
			& 3  & 4.54(-13) & 3.40(-12) & 1.17(-12) & 4.55(-13) & 3.40(-12) & 1.17(-12) & 5.21(-13) & 3.90(-12) & 1.34(-12) \\
			& 5  & 3.86(-08) & 1.99(-07) & 6.01(-08) & 3.86(-08) & 1.99(-07) & 6.01(-08) & 3.94(-08) & 2.04(-07) & 6.14(-08) \\
			& 10 & 1.62(-03) & 7.36(-03) & 6.37(-04) & 1.62(-03) & 7.36(-03) & 6.37(-04) & 1.63(-03) & 7.38(-03) & 6.38(-04) \\
			& 30 & 4.51(+02) & 9.91(+02) & 1.62(+01) & 4.51(+02) & 9.91(+02) & 1.62(+01) & 4.51(+02) & 9.93(+02) & 1.62(+01) \\
			\\{$^{78}$Ga} & 1  & 1.40(-47) & 1.40(-47) & 1.15(-39) & 2.37(-47) & 2.37(-47) & 1.95(-39) & 1.69(-45) & 1.69(-45) & 1.38(-37) \\
			& 2  & 3.48(-24) & 3.48(-24) & 4.81(-21) & 3.51(-24) & 3.50(-24) & 4.85(-21) & 7.33(-24) & 7.33(-24) & 1.01(-20) \\
			& 3  & 4.26(-16) & 4.22(-16) & 1.25(-14) & 4.26(-16) & 4.23(-16) & 1.25(-14) & 4.88(-16) & 4.84(-16) & 1.43(-14) \\
			& 5  & 2.65(-09) & 2.57(-09) & 3.53(-09) & 2.65(-09) & 2.58(-09) & 3.53(-09) & 2.71(-09) & 2.63(-09) & 3.61(-09) \\
			& 10 & 1.28(-03) & 1.19(-03) & 1.91(-04) & 1.28(-03) & 1.19(-03) & 1.91(-04) & 1.28(-03) & 1.19(-03) & 1.91(-04) \\
			& 30 & 4.10(+02) & 3.67(+02) & 1.47(+01) & 4.10(+02) & 3.68(+02) & 1.47(+01) & 4.10(+02) & 3.68(+02) & 1.47(+01) \\
			\\{$^{77}$Ge} & 1  & 1.02(-34) & 4.00(-34) & 1.57(-35) & 1.73(-34) & 6.76(-34) & 2.64(-35) & 1.24(-32) & 4.83(-32) & 1.88(-33) \\
			& 2  & 2.42(-19) & 1.74(-18) & 1.98(-19) & 2.43(-19) & 1.76(-18) & 2.00(-19) & 5.09(-19) & 3.67(-18) & 4.17(-19) \\
			& 3  & 5.92(-14) & 5.30(-13) & 1.05(-13) & 5.93(-14) & 5.30(-13) & 1.05(-13) & 6.79(-14) & 6.07(-13) & 1.21(-13) \\
			& 5  & 3.82(-09) & 2.88(-08) & 1.05(-08) & 3.82(-09) & 2.88(-08) & 1.05(-08) & 3.90(-09) & 2.94(-08) & 1.07(-08) \\
			& 10 & 8.34(-04) & 1.58(-03) & 2.94(-04) & 8.34(-04) & 1.58(-03) & 2.94(-04) & 8.38(-04) & 1.58(-03) & 2.94(-04) \\
			& 30 & 3.48(+02) & 4.43(+02) & 1.49(+01) & 3.48(+02) & 4.43(+02) & 1.49(+01) & 3.48(+02) & 4.43(+02) & 1.49(+01) \\
			\\{$^{71}$Cu} & 1  & 4.88(-44) & 7.66(-41) & 2.23(-42) & 8.24(-44) & 1.29(-40) & 3.77(-42) & 5.89(-42) & 9.25(-39) & 2.68(-40) \\
			& 2  & 5.73(-23) & 3.43(-21) & 9.62(-23) & 5.78(-23) & 3.46(-21) & 9.71(-23) & 1.21(-22) & 7.23(-21) & 2.02(-22) \\
			& 3  & 1.74(-15) & 2.36(-14) & 7.13(-16) & 1.75(-15) & 2.36(-14) & 7.14(-16) & 2.00(-15) & 2.70(-14) & 8.18(-16) \\
			& 5  & 3.94(-09) & 1.63(-08) & 5.55(-10) & 3.94(-09) & 1.63(-08) & 5.55(-10) & 4.03(-09) & 1.66(-08) & 5.66(-10) \\
			& 10 & 1.14(-03) & 3.14(-03) & 7.05(-05) & 1.14(-03) & 3.15(-03) & 7.05(-05) & 1.15(-03) & 3.16(-03) & 7.06(-05) \\
			& 30 & 4.37(+02) & 1.29(+03) & 1.01(+01) & 4.38(+02) & 1.29(+03) & 1.01(+01) & 4.38(+02) & 1.29(+03) & 1.01(+01) \\
			\\{$^{73}$Ga} & 1  & 9.12(-28) & 9.18(-29) & 2.65(-30) & 1.54(-27) & 1.55(-28) & 4.47(-30) & 1.10(-25) & 1.11(-26) & 3.18(-28) \\
			& 2  & 4.60(-15) & 1.03(-15) & 6.15(-17) & 4.63(-15) & 1.04(-15) & 6.19(-17) & 9.68(-15) & 2.16(-15) & 1.29(-16) \\
			& 3  & 1.44(-10) & 4.08(-11) & 4.28(-12) & 1.44(-10) & 4.08(-11) & 4.29(-12) & 1.65(-10) & 4.68(-11) & 4.91(-12) \\
			& 5  & 1.08(-06) & 3.99(-07) & 8.77(-08) & 1.08(-06) & 3.99(-07) & 8.77(-08) & 1.10(-06) & 4.08(-07) & 8.95(-08) \\
			& 10 & 5.81(-03) & 3.23(-03) & 7.74(-04) & 5.81(-03) & 3.24(-03) & 7.74(-04) & 5.82(-03) & 3.24(-03) & 7.76(-04) \\
			& 30 & 5.97(+02) & 2.14(+02) & 1.94(+01) & 5.97(+02) & 2.14(+02) & 1.94(+01) & 5.97(+02) & 2.14(+02) & 1.94(+01) \\
			\\\hline
	\end{tabular}}
\end{table*}

\begin{table*}[]
	\caption{Same as Table~\ref{T3ec} but for higher values of  density.}
	\label{T4ec}
	\centering
\resizebox{\textwidth}{!}{ 
		\begin{tabular}{c|c|ccc|ccc|ccc|}
			\hline
			\multicolumn{5}{l}{} & \multicolumn{1}{l}{$\lambda^{ec}$} & \multicolumn{5}{l}{}        \\
			\cline{1-11} \multicolumn{1}{l}{} &  & \multicolumn{3}{c}{$\rho$$\it Y_{e}$ = $10^7$ }        & \multicolumn{3}{c}{$\rho$$\it Y_{e}$ = $10^9$ }& \multicolumn{3}{c}{$\rho$$\it Y_{e}$ = $10^{11}$ }         \\
			\cline{3-11} \multicolumn{1}{l}{Nuclei} &       $T_9$                     & pn-QRPA-21& pn-QRPA-04    & IPM-03      & pn-QRPA-21 &pn-QRPA-04  & IPM-03      & pn-QRPA-21&pn-QRPA-04   & IPM-03 \\
			\hline
			\\{$^{75}$Ga} & 1  & 5.83(-31) & 2.19(-31) & 1.32(-32) & 2.61(-12) & 1.11(-12) & 7.78(-13) & 3.48(+04) & 3.48(+04) & 5.56(+03) \\
			& 2  & 7.67(-17) & 4.65(-17) & 4.11(-18) & 2.08(-07) & 1.42(-07) & 4.62(-08) & 3.57(+04) & 3.48(+04) & 5.57(+03) \\
			& 3  & 6.59(-12) & 4.46(-12) & 5.20(-13) & 1.91(-05) & 1.42(-05) & 3.92(-06) & 3.78(+04) & 3.49(+04) & 5.50(+03) \\
			& 5  & 9.55(-08) & 6.87(-08) & 1.33(-08) & 1.45(-03) & 1.10(-03) & 3.27(-04) & 4.21(+04) & 3.57(+04) & 5.53(+03) \\
			& 10 & 1.87(-03) & 1.31(-03) & 1.90(-04) & 3.18(-01) & 2.28(-01) & 3.51(-02) & 7.06(+04) & 5.35(+04) & 5.70(+03) \\
			& 30 & 5.14(+02) & 1.73(+02) & 1.23(+01) & 1.09(+03) & 3.68(+02) & 2.54(+01) & 8.75(+05) & 3.10(+05) & 1.07(+04) \\
			\\{$^{69}$Cu} & 1  & 1.50(-29) & 1.25(-27) & 6.28(-28) & 8.87(-10) & 1.44(-08) & 9.20(-10) & 2.84(+04) & 2.81(+04) & 6.03(+03) \\
			& 2  & 5.16(-16) & 6.49(-15) & 2.47(-15) & 3.89(-06) & 8.45(-06) & 1.90(-06) & 2.86(+04) & 3.01(+04) & 6.03(+03) \\
			& 3  & 2.37(-11) & 1.77(-10) & 6.10(-11) & 1.11(-04) & 1.43(-04) & 7.24(-05) & 2.91(+04) & 3.33(+04) & 5.97(+03) \\
			& 5  & 1.95(-07) & 1.00(-06) & 3.05(-07) & 3.32(-03) & 6.18(-03) & 3.48(-03) & 3.07(+04) & 3.95(+04) & 6.03(+03) \\
			& 10 & 2.03(-03) & 9.25(-03) & 8.00(-04) & 3.48(-01) & 1.46(+00) & 1.30(-01) & 5.09(+04) & 8.81(+04) & 6.50(+03) \\
			& 30 & 4.55(+02) & 1.00(+03) & 1.63(+01) & 9.62(+02) & 2.11(+03) & 3.37(+01) & 6.98(+05) & 1.35(+06) & 1.34(+04) \\
			\\{$^{78}$Ga} & 1  & 1.56(-41) & 1.56(-41) & 1.27(-33) & 1.70(-21) & 1.70(-21) & 8.71(-14) & 3.90(+04) & 3.89(+04) & 4.90(+03) \\
			& 2  & 2.50(-21) & 2.49(-21) & 3.44(-18) & 3.53(-11) & 3.53(-11) & 2.59(-08) & 4.16(+04) & 4.16(+04) & 4.95(+03) \\
			& 3  & 2.21(-14) & 2.20(-14) & 6.49(-13) & 1.82(-07) & 1.81(-07) & 3.26(-06) & 4.75(+04) & 4.72(+04) & 4.91(+03) \\
			& 5  & 1.34(-08) & 1.30(-08) & 1.79(-08) & 3.35(-04) & 3.26(-04) & 3.42(-04) & 6.03(+04) & 5.87(+04) & 5.02(+03) \\
			& 10 & 1.60(-03) & 1.50(-03) & 2.39(-04) & 3.00(-01) & 2.81(-01) & 4.21(-02) & 9.62(+04) & 9.04(+04) & 5.56(+03) \\
			& 30 & 4.13(+02) & 3.71(+02) & 1.48(+01) & 8.77(+02) & 7.87(+02) & 3.04(+01) & 7.64(+05) & 6.85(+05) & 1.30(+04) \\
			\\{$^{77}$Ge} & 1  & 1.14(-28) & 4.47(-28) & 1.73(-29) & 8.51(-09) & 3.89(-09) & 1.80(-09) & 3.13(+04) & 4.51(+04) & 6.38(+03) \\
			& 2  & 1.73(-16) & 1.25(-15) & 1.42(-16) & 1.85(-06) & 1.61(-06) & 1.97(-06) & 3.94(+04) & 6.19(+04) & 6.41(+03) \\
			& 3  & 3.08(-12) & 2.75(-11) & 5.46(-12) & 2.05(-05) & 3.51(-05) & 4.43(-05) & 4.56(+04) & 6.97(+04) & 6.32(+03) \\
			& 5  & 1.92(-08) & 1.45(-07) & 5.32(-08) & 4.20(-04) & 1.53(-03) & 1.31(-03) & 5.20(+04) & 7.59(+04) & 6.37(+03) \\
			& 10 & 1.05(-03) & 1.98(-03) & 3.69(-04) & 1.91(-01) & 3.39(-01) & 6.81(-02) & 7.36(+04) & 9.89(+04) & 6.56(+03) \\
			& 30 & 3.51(+02) & 4.47(+02) & 1.50(+01) & 7.45(+02) & 9.46(+02) & 3.08(+01) & 6.55(+05) & 8.09(+05) & 1.23(+04) \\
			\\{$^{71}$Cu} & 1  & 5.45(-38) & 8.55(-35) & 2.47(-36) & 8.07(-19) & 3.78(-17) & 2.62(-16) & 3.24(+04) & 3.27(+04) & 4.75(+03) \\
			& 2  & 4.12(-20) & 2.47(-18) & 6.89(-20) & 3.77(-10) & 2.10(-09) & 9.62(-10) & 3.26(+04) & 3.49(+04) & 4.80(+03) \\
			& 3  & 9.06(-14) & 1.23(-12) & 3.71(-14) & 6.46(-07) & 2.14(-06) & 3.03(-07) & 3.30(+04) & 3.85(+04) & 4.73(+03) \\
			& 5  & 2.00(-08) & 8.22(-08) & 2.81(-09) & 4.81(-04) & 1.22(-03) & 6.98(-05) & 3.48(+04) & 4.54(+04) & 4.78(+03) \\
			& 10 & 1.44(-03) & 3.94(-03) & 8.85(-05) & 2.65(-01) & 6.90(-01) & 1.64(-02) & 5.75(+04) & 9.62(+04) & 4.93(+03) \\
			& 30 & 4.41(+02) & 1.30(+03) & 1.02(+01) & 9.33(+02) & 2.75(+03) & 2.09(+01) & 7.26(+05) & 2.00(+06) & 9.48(+03) \\
			\\{$^{73}$Ga} & 1  & 1.01(-21) & 1.02(-22) & 2.92(-24) & 8.75(-04) & 1.80(-04) & 1.30(-04) & 3.64(+04) & 3.54(+04) & 7.40(+03) \\
			& 2  & 3.29(-12) & 7.35(-13) & 4.40(-14) & 2.67(-03) & 3.56(-04) & 4.83(-04) & 3.74(+04) & 3.52(+04) & 7.40(+03) \\
			& 3  & 7.45(-09) & 2.11(-09) & 2.22(-10) & 8.18(-03) & 1.06(-03) & 1.60(-03) & 3.95(+04) & 3.52(+04) & 7.29(+03) \\
			& 5  & 5.42(-06) & 2.01(-06) & 4.44(-07) & 4.25(-02) & 1.15(-02) & 1.04(-02) & 4.43(+04) & 3.59(+04) & 7.33(+03) \\
			& 10 & 7.28(-03) & 4.06(-03) & 9.73(-04) & 1.03(+00) & 5.85(-01) & 1.76(-01) & 7.69(+04) & 5.62(+04) & 7.52(+03) \\
			& 30 & 6.03(+02) & 2.16(+02) & 1.95(+01) & 1.27(+03) & 4.57(+02) & 4.01(+01) & 9.14(+05) & 3.52(+05) & 1.46(+04) \\
			\\\hline
	\end{tabular}}
\end{table*}

\begin{table*}[]
	\caption{The \textit{bd} rates ($\lambda^{bd}$) for $^{67}$Ni, $^{69}$Cu, $^{67}$Co, $^{66}$Co, $^{71}$Cu and $^{77}$Ga. The previous pn-QRPA rates~\cite{Nab04} and IPM calculation~\cite{Pru03} are also shown for comparison. The units of $T_9$, $\rho$$\it Y_{e}$ and $\lambda^{bd}$ are GK, g/cm$^3$ and $s^{-1}$, respectively. The nuclei are ordered on the basis of ranking shown in Table~\ref{T1}. Number in parenthesis denotes exponents. In the table, $1.00(-100)$ means that the rate is smaller than 10$^{-100} (s^{-1})$.}
	\label{T1bd}
	\centering
\resizebox{\textwidth}{!}{ 
		\begin{tabular}{c|c|ccc|ccc|ccc|}
			\hline
			\multicolumn{5}{l}{} & \multicolumn{1}{l}{$\lambda^{bd}$} & \multicolumn{5}{l}{}        \\
			\cline{1-11} \multicolumn{1}{l}{} &  & \multicolumn{3}{c}{$\rho$$\it Y_{e}$ = $10^1$ }        & \multicolumn{3}{c}{$\rho$$\it Y_{e}$ = $10^3$ }& \multicolumn{3}{c}{$\rho$$\it Y_{e}$ = $10^5$ }         \\
			\cline{3-11} \multicolumn{1}{l}{Nuclei} &       $T_9$                     & pn-QRPA-21& pn-QRPA-04    & IPM-03      & pn-QRPA-21 &pn-QRPA-04  & IPM-03      & pn-QRPA-21&pn-QRPA-04   & IPM-03 \\
			\hline
			\\ {$^{67}$Ni} & 1 & 7.28(-02) & 5.68(-02) & 3.59(-02) & 7.28(-02) & 5.68(-02) & 3.59(-02) & 7.26(-02) & 5.65(-02) & 3.58(-02) \\
			& 2 & 8.55(-02) & 6.21(-02) & 4.94(-02) & 8.55(-02) & 6.21(-02) & 4.94(-02) & 8.53(-02) & 6.19(-02) & 4.94(-02) \\
			&3 & 8.89(-02) & 6.55(-02) & 6.81(-02) & 8.89(-02) & 6.55(-02) & 6.81(-02) & 8.89(-02) & 6.53(-02) & 6.81(-02) \\
			&5 & 8.81(-02) & 7.52(-02) & 1.49(-01) & 8.81(-02) & 7.52(-02) & 1.49(-01) & 8.79(-02) & 7.52(-02) & 1.49(-01) \\
			& 10 & 1.98(-01) & 2.37(-01) & 7.01(-01) & 1.98(-01) & 2.37(-01) & 7.01(-01) & 1.98(-01) & 2.37(-01) & 7.01(-01) \\
			& 30 & 1.35(+01) & 1.12(+01) & 9.82(-01) & 1.35(+01) & 1.12(+01) & 9.82(-01) & 1.35(+01) & 1.12(+01) & 9.82(-01) \\
			\\{$^{69}$Cu} & 1 & 1.72(-01) & 3.44(-03) & 4.42(-03) & 1.72(-01) & 3.44(-03) & 4.42(-03) & 1.71(-01) & 3.42(-03) & 4.38(-03) \\
			& 2 & 1.57(-01) & 1.60(-02) & 6.59(-03) & 1.57(-01) & 1.60(-02) & 6.59(-03) & 1.57(-01) & 1.60(-02) & 6.58(-03) \\
			&3 & 1.55(-01) & 3.83(-02) & 1.44(-02) & 1.55(-01) & 3.83(-02) & 1.44(-02) & 1.55(-01) & 3.82(-02) & 1.43(-02) \\
			&5 & 1.61(-01) & 8.63(-02) & 7.31(-02) & 1.61(-01) & 8.63(-02) & 7.31(-02) & 1.60(-01) & 8.63(-02) & 7.29(-02) \\
			& 10 & 2.19(-01) & 3.64(-01) & 2.77(-01) & 2.19(-01) & 3.64(-01) & 2.77(-01) & 2.19(-01) & 3.63(-01) & 2.77(-01) \\
			& 30 & 2.66(+00) & 5.11(+00) & 4.53(-01) & 2.66(+00) & 5.11(+00) & 4.53(-01) & 2.66(+00) & 5.11(+00) & 4.53(-01) \\
			\\ {$^{67}$Co} & 1 & 5.82(+00) & 1.46(+01) & 1.70(+00) & 5.82(+00) & 1.46(+01) & 1.70(+00) & 5.81(+00) & 1.46(+01) & 1.69(+00) \\
			& 2 & 6.53(+00) & 2.01(+01) & 1.76(+00) & 6.53(+00) & 2.01(+01) & 1.76(+00) & 6.53(+00) & 2.01(+01) & 1.76(+00) \\
			&3 & 7.18(+00) & 2.43(+01) & 1.85(+00) & 7.18(+00) & 2.43(+01) & 1.85(+00) & 7.18(+00) & 2.43(+01) & 1.85(+00) \\
			&5 & 8.02(+00) & 2.92(+01) & 2.13(+00) & 8.02(+00) & 2.92(+01) & 2.13(+00) & 8.02(+00) & 2.92(+01) & 2.13(+00) \\
			& 10 & 1.29(+01) & 8.85(+01) & 3.33(+00) & 1.29(+01) & 8.85(+01) & 3.33(+00) & 1.29(+01) & 8.85(+01) & 3.33(+00) \\
			& 30 & 9.53(+01) & 1.42(+03) & 2.26(+01) & 9.53(+01) & 1.42(+03) & 2.26(+01) & 9.53(+01) & 1.42(+03) & 2.26(+01) \\
			\\ {$^{66}$Co} & 1 & 3.01(+00) & 1.15(+01) & 2.94(+00) & 3.01(+00) & 1.15(+01) & 2.94(+00) & 3.01(+00) & 1.15(+01) & 2.94(+00) \\
			& 2 & 4.71(+00) & 1.48(+01) & 2.97(+00) & 4.71(+00) & 1.48(+01) & 2.97(+00) & 4.71(+00) & 1.48(+01) & 2.97(+00) \\
			&3 & 5.74(+00) & 1.69(+01) & 2.98(+00) & 5.74(+00) & 1.69(+01) & 2.98(+00) & 5.74(+00) & 1.69(+01) & 2.98(+00) \\
			&5 & 6.71(+00) & 1.97(+01) & 2.99(+00) & 6.71(+00) & 1.97(+01) & 2.99(+00) & 6.71(+00) & 1.97(+01) & 2.99(+00) \\
			& 10 & 8.32(+00) & 3.31(+01) & 3.07(+00) & 8.32(+00) & 3.31(+01) & 3.07(+00) & 8.32(+00) & 3.31(+01) & 3.07(+00) \\
			& 30 & 3.79(+01) & 3.73(+02) & 4.29(+01) & 3.79(+01) & 3.73(+02) & 4.29(+01) & 3.79(+01) & 3.73(+02) & 4.29(+01) \\
			\\ {$^{71}$Cu} & 1 & 1.44(+00) & 1.16(+00) & 4.16(-02) & 1.44(+00) & 1.16(+00) & 4.16(-02) & 1.43(+00) & 1.16(+00) & 4.15(-02) \\
			& 2 & 1.31(+00) & 1.20(+00) & 7.50(-02) & 1.31(+00) & 1.20(+00) & 7.50(-02) & 1.30(+00) & 1.20(+00) & 7.50(-02) \\
			&3 & 1.25(+00) & 1.28(+00) & 1.35(-01) & 1.25(+00) & 1.28(+00) & 1.35(-01) & 1.25(+00) & 1.28(+00) & 1.34(-01) \\
			&5 & 1.24(+00) & 1.44(+00) & 3.22(-01) & 1.24(+00) & 1.44(+00) & 3.22(-01) & 1.24(+00) & 1.44(+00) & 3.22(-01) \\
			& 10 & 1.64(+00) & 3.76(+00) & 1.04(+00) & 1.64(+00) & 3.76(+00) & 1.04(+00) & 1.64(+00) & 3.76(+00) & 1.04(+00) \\
			& 30 & 1.35(+01) & 1.09(+02) & 2.24(+00) & 1.35(+01) & 1.09(+02) & 2.24(+00) & 1.35(+01) & 1.09(+02) & 2.24(+00) \\
			\\ {$^{77}$Ga} & 1 & 3.13(-01) & 3.58(-01) & 7.06(-02) & 3.13(-01) & 3.58(-01) & 7.06(-02) & 3.13(-01) & 3.57(-01) & 7.06(-02) \\
			& 2 & 3.18(-01) & 3.65(-01) & 1.20(-01) & 3.18(-01) & 3.65(-01) & 1.20(-01) & 3.18(-01) & 3.65(-01) & 1.20(-01) \\
			&3 & 3.60(-01) & 4.14(-01) & 1.94(-01) & 3.60(-01) & 4.14(-01) & 1.94(-01) & 3.60(-01) & 4.13(-01) & 1.94(-01) \\
			&5 & 5.81(-01) & 6.65(-01) & 4.26(-01) & 5.81(-01) & 6.65(-01) & 4.26(-01) & 5.81(-01) & 6.65(-01) & 4.26(-01) \\
			& 10 & 1.86(+00) & 2.12(+00) & 1.45(+00) & 1.86(+00) & 2.12(+00) & 1.45(+00) & 1.86(+00) & 2.12(+00) & 1.45(+00) \\
			& 30 & 4.86(+01) & 4.20(+01) & 4.20(+00) & 4.86(+01) & 4.20(+01) & 4.20(+00) & 4.86(+01) & 4.20(+01) & 4.20(+00)\\
			\\	\hline
	\end{tabular}}
\end{table*}

\begin{table*}[]
	\caption{Same as Table~\ref{T1bd} but for higher values of  density.}
	\label{T2bd}
	\centering
\resizebox{\textwidth}{!}{ 
		\begin{tabular}{c|c|ccc|ccc|ccc|}
			\hline
			\multicolumn{5}{l}{} & \multicolumn{1}{l}{$\lambda^{bd}$} & \multicolumn{5}{l}{}        \\
			\cline{1-11} \multicolumn{1}{l}{} &  & \multicolumn{3}{c}{$\rho$$\it Y_{e}$ = $10^7$ }        & \multicolumn{3}{c}{$\rho$$\it Y_{e}$ = $10^9$ }& \multicolumn{3}{c}{$\rho$$\it Y_{e}$ = $10^{11}$ }         \\
			\cline{3-11} \multicolumn{1}{l}{Nuclei} &       $T_9$                     & pn-QRPA-21& pn-QRPA-04    & IPM-03      & pn-QRPA-21 &pn-QRPA-04  & IPM-03      & pn-QRPA-21&pn-QRPA-04   & IPM-03 \\
			\hline 
			\\{$^{67}$Ni} & 1 & 6.17(-02) & 4.17(-02) & 3.05(-02) & 1.03(-10) & 4.99(-11) & 5.74(-10) & 1.00(-100) & 1.00(-100) & 1.00(-100) \\
			& 2 & 7.35(-02) & 4.67(-02) & 4.31(-02) & 5.12(-07) & 2.74(-07) & 3.70(-06) & 1.71(-53) & 6.49(-54) & 1.00(-100) \\
			&3 & 7.80(-02) & 5.14(-02) & 6.12(-02) & 1.43(-05) & 9.35(-06) & 1.23(-04) & 2.47(-36) & 1.02(-36) & 1.00(-100) \\
			&5 & 8.11(-02) & 6.56(-02) & 1.41(-01) & 3.35(-04) & 2.97(-04) & 3.71(-03) & 2.17(-22) & 1.07(-22) & 5.64(-22)  \\
			& 10 & 1.95(-01) & 2.32(-01) & 6.92(-01) & 3.49(-02) & 3.43(-02) & 1.22(-01) & 2.82(-11) & 2.50(-11) & 5.38(-11)  \\
			& 30 & 1.35(+01) & 1.12(+01) & 9.79(-01) & 1.10(+01) & 8.97(+00) & 7.80(-01) & 1.16(-02) & 8.85(-03) & 6.70(-04) \\
			\\{$^{69}$Cu} & 1 & 1.29(-01) & 2.51(-03) & 2.68(-03) & 3.60(-14) & 1.19(-14) & 8.99(-15) & 1.00(-100) & 1.00(-100) & 1.00(-100) \\
			& 2 & 1.22(-01) & 1.27(-02) & 4.66(-03) & 2.52(-08) & 1.35(-08) & 1.25(-08) & 1.37(-55) & 1.03(-55) & 1.00(-100) \\
			&3 & 1.26(-01) & 3.18(-02) & 1.17(-02) & 3.31(-06) & 2.38(-06) & 3.50(-06) & 9.66(-38) & 9.12(-38) & 1.00(-100) \\
			&5 & 1.43(-01) & 7.78(-02) & 6.75(-02) & 2.32(-04) & 2.32(-04) & 6.73(-04) & 2.48(-23) & 3.06(-23) & 3.94(-23)  \\
			& 10 & 2.13(-01) & 3.53(-01) & 2.72(-01) & 1.17(-02) & 2.24(-02) & 3.31(-02) & 3.08(-12) & 6.75(-12) & 1.15(-11)  \\
			& 30 & 2.65(+00) & 5.09(+00) & 4.52(-01) & 2.01(+00) & 3.78(+00) & 3.52(-01) & 1.40(-03) & 2.46(-03) & 2.68(-04) \\
			\\ {$^{67}$Co} & 1 & 5.66(+00) & 1.39(+01) & 1.66(+00) & 1.78(+00) & 1.98(+00) & 4.44(-01) & 8.41(-80) & 7.53(-80) & 1.00(-100) \\
			& 2 & 6.37(+00) & 1.93(+01) & 1.72(+00) & 2.01(+00) & 3.05(+00) & 4.78(-01) & 4.35(-41) & 5.11(-41) & 1.00(-100) \\
			&3 & 7.01(+00) & 2.34(+01) & 1.81(+00) & 2.25(+00) & 4.05(+00) & 5.28(-01) & 5.85(-28) & 7.23(-28) & 3.90(-30) \\
			&5 & 7.89(+00) & 2.85(+01) & 2.10(+00) & 2.69(+00) & 6.00(+00) & 7.10(-01) & 3.18(-17) & 5.14(-17) & 3.16(-17) \\
			& 10 & 1.28(+01) & 8.79(+01) & 3.31(+00) & 5.52(+00) & 3.65(+01) & 1.63(+00) & 1.20(-08) & 7.83(-08) & 8.26(-09) \\
			& 30 & 9.51(+01) & 1.42(+03) & 2.26(+01) & 8.05(+01) & 1.21(+03) & 1.99(+01) & 1.14(-01) & 1.62(+00) & 3.92(-02)  \\
			\\ {$^{66}$Co} & 1 & 2.88(+00) & 1.08(+01) & 2.86(+00) & 3.65(-01) & 7.45(-01) & 6.07(-01) & 2.99(-88) & 1.04(-85) & 1.00(-100) \\
			& 2 & 4.53(+00) & 1.40(+01) & 2.89(+00) & 6.40(-01) & 1.04(+00) & 6.37(-01) & 4.02(-45) & 5.58(-44) & 1.00(-100) \\
			&3 & 5.55(+00) & 1.61(+01) & 2.91(+00) & 8.57(-01) & 1.40(+00) & 6.70(-01) & 1.52(-30) & 8.79(-30) & 1.02(-26) \\
			&5 & 6.56(+00) & 1.91(+01) & 2.93(+00) & 1.22(+00) & 2.28(+00) & 7.59(-01) & 1.11(-18) & 3.72(-18) & 1.03(-16) \\
			& 10 & 8.24(+00) & 3.27(+01) & 3.05(+00) & 2.53(+00) & 8.55(+00) & 1.17(+00) & 2.00(-09) & 7.57(-09) & 5.11(-09) \\
			& 30 & 3.78(+01) & 3.72(+02) & 4.29(+01) & 3.10(+01) & 3.05(+02) & 3.84(+01) & 3.21(-02) & 3.12(-01) & 1.03(-01)  \\
			\\ {$^{71}$Cu} & 1 & 1.29(+00) & 1.05(+00) & 3.82(-02) & 8.81(-06) & 4.23(-06) & 7.31(-06) & 4.61(-100) & 1.94(-99) & 1.00(-100) \\
			& 2 & 1.17(+00) & 1.09(+00) & 6.97(-02) & 5.08(-04) & 3.78(-04) & 2.84(-04) & 3.27(-51) & 7.57(-51) & 1.00(-100) \\
			&3 & 1.14(+00) & 1.17(+00) & 1.26(-01) & 3.01(-03) & 2.79(-03) & 2.23(-03) & 1.01(-34) & 2.13(-34) & 1.00(-100) \\
			&5 & 1.17(+00) & 1.36(+00) & 3.11(-01) & 1.91(-02) & 2.22(-02) & 2.22(-02) & 2.36(-21) & 5.01(-21) & 9.48(-21)  \\
			& 10 & 1.61(+00) & 3.69(+00) & 1.03(+00) & 1.86(-01) & 4.23(-01) & 2.57(-01) & 5.83(-11) & 1.74(-10) & 1.66(-10) \\
			& 30 & 1.35(+01) & 1.08(+02) & 2.24(+00) & 1.05(+01) & 8.26(+01) & 1.83(+00) & 8.32(-03) & 5.74(-02) & 1.82(-03)  \\
			\\ {$^{77}$Ga} & 1 & 2.78(-01) & 3.19(-01) & 6.67(-02) & 2.38(-06) & 9.95(-06) & 1.21(-03) & 4.20(-97) & 1.68(-96) & 1.00(-100) \\
			& 2 & 2.86(-01) & 3.28(-01) & 1.14(-01) & 3.78(-04) & 6.38(-04) & 3.85(-03) & 6.01(-50) & 1.29(-49) & 1.00(-100) \\
			&3 & 3.30(-01) & 3.80(-01) & 1.85(-01) & 3.73(-03) & 5.36(-03) & 1.13(-02) & 8.09(-34) & 1.39(-33) & 1.00(-100) \\
			&5 & 5.56(-01) & 6.37(-01) & 4.15(-01) & 3.26(-02) & 4.18(-02) & 5.53(-02) & 1.17(-20) & 1.66(-20) & 5.77(-20) \\
			& 10 & 1.84(+00) & 2.10(+00) & 1.43(+00) & 4.09(-01) & 4.89(-01) & 4.54(-01) & 2.38(-10) & 2.92(-10) & 4.30(-10) \\
			& 30 & 4.85(+01) & 4.20(+01) & 4.20(+00) & 3.94(+01) & 3.40(+01) & 3.48(+00) & 3.93(-02) & 3.24(-02) & 3.94(-03) \\
			\\	\hline
	\end{tabular}}
\end{table*}

\begin{table*}[]
	\caption{Same as Table~\ref{T1bd} but for $^{75}$Ga, $^{79}$Ga, $^{70}$Cu, $^{68}$Ni and $^{68}$Co.}
	\label{T3bd}
	\centering
	\resizebox{\textwidth}{!}{ 
		\begin{tabular}{c|c|ccc|ccc|ccc|}
			\hline
			\multicolumn{5}{l}{} & \multicolumn{1}{l}{$\lambda^{bd}$} & \multicolumn{5}{l}{}        \\
			\cline{1-11} \multicolumn{1}{l}{} &  & \multicolumn{3}{c}{$\rho$$\it Y_{e}$ = $10^1$ }        & \multicolumn{3}{c}{$\rho$$\it Y_{e}$ = $10^3$ }& \multicolumn{3}{c}{$\rho$$\it Y_{e}$ = $10^5$ }         \\
			\cline{3-11} \multicolumn{1}{l}{Nuclei} &       $T_9$                     & pn-QRPA-21& pn-QRPA-04    & IPM-03      & pn-QRPA-21 &pn-QRPA-04  & IPM-03      & pn-QRPA-21&pn-QRPA-04   & IPM-03 \\
			\hline
			\\{$^{75}$Ga} & 1 & 5.37(-02) & 5.36(-02) & 9.53(-03) & 5.37(-02) & 5.36(-02) & 9.53(-03) & 5.36(-02) & 5.33(-02) & 9.48(-03) \\
			& 2 & 5.53(-02) & 5.36(-02) & 2.42(-02) & 5.53(-02) & 5.36(-02) & 2.42(-02) & 5.52(-02) & 5.35(-02) & 2.42(-02) \\
			&3 & 6.68(-02) & 6.15(-02) & 4.72(-02) & 6.68(-02) & 6.15(-02) & 4.72(-02) & 6.67(-02) & 6.14(-02) & 4.71(-02) \\
			&5 & 1.26(-01) & 1.07(-01) & 1.25(-01) & 1.26(-01) & 1.07(-01) & 1.25(-01) & 1.26(-01) & 1.07(-01) & 1.25(-01) \\
			& 10 & 4.98(-01) & 3.66(-01) & 4.46(-01) & 4.98(-01) & 3.66(-01) & 4.46(-01) & 4.98(-01) & 3.66(-01) & 4.46(-01) \\
			& 30 & 1.37(+01) & 3.22(+00) & 9.66(-01) & 1.37(+01) & 3.22(+00) & 9.66(-01) & 1.37(+01) & 3.22(+00) & 9.66(-01) \\
			\\ {$^{79}$Ga} & 1 & 9.44(-01) & 9.73(-01) & 3.09(-01) & 9.44(-01) & 9.73(-01) & 3.09(-01) & 9.42(-01) & 9.71(-01) & 3.08(-01) \\
			& 2 & 9.23(-01) & 9.79(-01) & 4.10(-01) & 9.23(-01) & 9.79(-01) & 4.10(-01) & 9.20(-01) & 9.79(-01) & 4.10(-01) \\
			&3 & 9.02(-01) & 1.03(+00) & 5.94(-01) & 9.02(-01) & 1.03(+00) & 5.94(-01) & 9.02(-01) & 1.03(+00) & 5.94(-01) \\
			&5 & 1.07(+00) & 1.41(+00) & 1.15(+00) & 1.07(+00) & 1.41(+00) & 1.15(+00) & 1.07(+00) & 1.41(+00) & 1.15(+00) \\
			& 10 & 3.27(+00) & 5.01(+00) & 4.56(+00) & 3.27(+00) & 5.01(+00) & 4.56(+00) & 3.27(+00) & 5.01(+00) & 4.56(+00) \\
			& 30 & 1.17(+02) & 1.15(+02) & 1.13(+01) & 1.17(+02) & 1.15(+02) & 1.13(+01) & 1.17(+02) & 1.15(+02) & 1.13(+01) \\
			\\{$^{70}$Cu} & 1 & 6.08(-01) & 4.76(-01) & 1.09(-01) & 6.08(-01) & 4.76(-01) & 1.09(-01) & 6.07(-01) & 4.75(-01) & 1.09(-01) \\
			& 2 & 7.13(-01) & 5.55(-01) & 8.75(-02) & 7.13(-01) & 5.55(-01) & 8.75(-02) & 7.11(-01) & 5.55(-01) & 8.75(-02) \\
			&3 & 8.09(-01) & 6.97(-01) & 8.22(-02) & 8.09(-01) & 6.97(-01) & 8.22(-02) & 8.09(-01) & 6.95(-01) & 8.20(-02) \\
			&5 & 1.01(+00) & 1.10(+00) & 8.67(-02) & 1.01(+00) & 1.10(+00) & 8.67(-02) & 1.01(+00) & 1.10(+00) & 8.67(-02) \\
			& 10 & 1.48(+00) & 2.48(+00) & 1.97(-01) & 1.48(+00) & 2.48(+00) & 1.97(-01) & 1.48(+00) & 2.48(+00) & 1.97(-01) \\
			& 30 & 8.18(+00) & 1.23(+01) & 8.39(+00) & 8.18(+00) & 1.23(+01) & 8.39(+00) & 8.18(+00) & 1.23(+01) & 8.39(+00) \\
			\\{$^{68}$Ni} & 1 & 3.05(-03) & 9.53(-04) & 2.40(-02) & 3.05(-03) & 9.53(-04) & 2.40(-02) & 3.01(-03) & 9.31(-04) & 2.39(-02) \\
			& 2 & 3.04(-03) & 9.44(-04) & 2.42(-02) & 3.04(-03) & 9.44(-04) & 2.42(-02) & 3.02(-03) & 9.35(-04) & 2.41(-02) \\
			&3 & 2.98(-03) & 9.23(-04) & 2.99(-02) & 2.98(-03) & 9.23(-04) & 2.99(-02) & 2.96(-03) & 9.18(-04) & 2.98(-02) \\
			&5 & 3.16(-03) & 1.70(-03) & 2.09(-01) & 3.16(-03) & 1.70(-03) & 2.09(-01) & 3.15(-03) & 1.70(-03) & 2.09(-01) \\
			& 10 & 1.19(-01) & 1.74(-01) & 2.14(+00) & 1.19(-01) & 1.74(-01) & 2.14(+00) & 1.19(-01) & 1.74(-01) & 2.14(+00) \\
			& 30 & 1.13(+01) & 7.05(+00) & 2.64(-01) & 1.13(+01) & 7.05(+00) & 2.64(-01) & 1.13(+01) & 7.05(+00) & 2.64(-01) \\
			\\ {$^{68}$Co} & 1 & 8.34(+00) & 2.34(+01) & 4.25(+00) & 8.34(+00) & 2.34(+01) & 4.25(+00) & 8.34(+00) & 2.34(+01) & 4.25(+00) \\
			& 2 & 1.53(+01) & 4.06(+01) & 4.62(+00) & 1.53(+01) & 4.06(+01) & 4.62(+00) & 1.53(+01) & 4.06(+01) & 4.62(+00) \\
			&3 & 2.00(+01) & 5.20(+01) & 4.76(+00) & 2.00(+01) & 5.20(+01) & 4.76(+00) & 2.00(+01) & 5.20(+01) & 4.76(+00) \\
			&5 & 2.40(+01) & 6.22(+01) & 4.83(+00) & 2.40(+01) & 6.22(+01) & 4.83(+00) & 2.40(+01) & 6.22(+01) & 4.83(+00) \\
			& 10 & 2.92(+01) & 1.36(+02) & 6.85(+00) & 2.92(+01) & 1.36(+02) & 6.85(+00) & 2.92(+01) & 1.36(+02) & 6.85(+00) \\
			& 30 & 1.26(+02) & 2.57(+03) & 8.43(+01) & 1.26(+02) & 2.57(+03) & 8.43(+01) & 1.26(+02) & 2.57(+03) & 8.43(+01)\\
			\\	\hline
	\end{tabular}}
\end{table*}

\begin{table*}[]
	\caption{Same as Table~\ref{T3bd} but for higher values of  density.}
	\label{T4bd}
	\centering
	\resizebox{\textwidth}{!}{ 
		\begin{tabular}{c|c|ccc|ccc|ccc|}
			\hline
			\multicolumn{5}{l}{} & \multicolumn{1}{l}{$\lambda^{bd}$} & \multicolumn{5}{l}{}        \\
			\cline{1-11} \multicolumn{1}{l}{} &  & \multicolumn{3}{c}{$\rho$$\it Y_{e}$ = $10^7$ }        & \multicolumn{3}{c}{$\rho$$\it Y_{e}$ = $10^9$ }& \multicolumn{3}{c}{$\rho$$\it Y_{e}$ = $10^{11}$ }         \\
			\cline{3-11} \multicolumn{1}{l}{Nuclei} &       $T_9$                     & pn-QRPA-21& pn-QRPA-04    & IPM-03      & pn-QRPA-21 &pn-QRPA-04  & IPM-03      & pn-QRPA-21&pn-QRPA-04   & IPM-03 \\
			\hline
			\\ {$^{75}$Ga} & 1 & 4.21(-02) & 4.20(-02) & 7.94(-03) & 1.51(-11) & 1.59(-11) & 4.44(-11) & 1.00(-100) & 1.00(-100) & 1.00(-100) \\
			& 2 & 4.46(-02) & 4.32(-02) & 2.08(-02) & 6.30(-07) & 6.24(-07) & 1.24(-06) & 4.95(-54) & 4.86(-54) & 1.00(-100) \\
			&3 & 5.68(-02) & 5.24(-02) & 4.21(-02) & 3.59(-05) & 3.33(-05) & 6.50(-05) & 1.50(-36) & 1.37(-36) & 1.00(-100) \\
			&5 & 1.17(-01) & 9.95(-02) & 1.18(-01) & 1.46(-03) & 1.23(-03) & 2.53(-03) & 2.19(-22) & 1.82(-22) & 3.19(-22)  \\
			& 10 & 4.89(-01) & 3.60(-01) & 4.39(-01) & 6.32(-02) & 4.54(-02) & 7.08(-02) & 2.34(-11) & 1.60(-11) & 2.95(-11)  \\
			& 30 & 1.37(+01) & 3.21(+00) & 9.64(-01) & 1.08(+01) & 2.50(+00) & 7.62(-01) & 9.06(-03) & 1.90(-03) & 6.38(-04) \\
			\\ {$^{79}$Ga} & 1 & 8.39(-01) & 8.63(-01) & 2.93(-01) & 7.87(-03) & 9.33(-03) & 2.96(-02) & 4.13(-88) & 5.33(-88) & 1.00(-100) \\
			& 2 & 8.28(-01) & 8.79(-01) & 3.94(-01) & 1.26(-02) & 1.46(-02) & 5.55(-02) & 1.13(-45) & 1.39(-45) & 1.00(-100) \\
			&3 & 8.24(-01) & 9.38(-01) & 5.75(-01) & 2.46(-02) & 3.03(-02) & 1.09(-01) & 2.88(-31) & 3.72(-31) & 1.00(-100) \\
			&5 & 1.02(+00) & 1.35(+00) & 1.13(+00) & 1.04(-01) & 1.46(-01) & 3.16(-01) & 2.54(-19) & 3.60(-19) & 3.13(-18) \\
			& 10 & 3.23(+00) & 4.97(+00) & 4.54(+00) & 1.02(+00) & 1.57(+00) & 2.06(+00) & 1.22(-09) & 1.82(-09) & 5.27(-09) \\
			& 30 & 1.17(+02) & 1.14(+02) & 1.13(+01) & 9.73(+01) & 9.51(+01) & 9.71(+00) & 1.17(-01) & 1.10(-01) & 1.47(-02)  \\
			\\ {$^{70}$Cu} & 1 & 5.26(-01) & 4.09(-01) & 1.04(-01) & 6.35(-09) & 7.52(-03) & 8.28(-03) & 1.00(-100) & 6.85(-91) & 1.00(-100) \\
			& 2 & 6.24(-01) & 4.74(-01) & 8.26(-02) & 1.71(-05) & 5.83(-03) & 7.14(-03) & 1.02(-52) & 9.31(-48) & 1.00(-100) \\
			&3 & 7.23(-01) & 6.08(-01) & 7.80(-02) & 3.94(-04) & 5.51(-03) & 7.91(-03) & 1.28(-35) & 3.33(-33) & 1.00(-100) \\
			&5 & 9.44(-01) & 1.01(+00) & 8.39(-02) & 7.62(-03) & 1.26(-02) & 1.29(-02) & 9.25(-22) & 3.43(-21) & 5.01(-20) \\
			& 10 & 1.45(+00) & 2.42(+00) & 1.95(-01) & 1.41(-01) & 2.14(-01) & 7.26(-02) & 4.33(-11) & 7.43(-11) & 1.32(-10) \\
			& 30 & 8.17(+00) & 1.23(+01) & 8.39(+00) & 6.24(+00) & 9.35(+00) & 7.16(+00) & 4.43(-03) & 6.64(-03) & 1.01(-02)  \\
			\\ {$^{68}$Ni} & 1 & 1.19(-03) & 2.62(-04) & 1.45(-02) & 5.26(-21) & 5.83(-21) & 7.78(-17) & 1.00(-100) & 1.00(-100) & 1.00(-100) \\
			& 2 & 1.41(-03) & 3.52(-04) & 1.56(-02) & 8.73(-12) & 1.31(-11) & 3.37(-09) & 3.96(-56) & 2.96(-56) & 1.00(-100) \\
			&3 & 1.69(-03) & 4.59(-04) & 2.22(-02) & 2.71(-08) & 3.50(-08) & 2.46(-06) & 1.13(-37) & 9.73(-38) & 1.00(-100) \\
			&5 & 2.47(-03) & 1.42(-03) & 1.93(-01) & 5.65(-05) & 6.04(-05) & 1.34(-03) & 1.12(-22) & 1.12(-22) & 3.10(-23)  \\
			& 10 & 1.18(-01) & 1.71(-01) & 2.09(+00) & 3.97(-02) & 4.43(-02) & 2.04(-01) & 4.79(-11) & 5.56(-11) & 6.22(-11)  \\
			& 30 & 1.12(+01) & 7.03(+00) & 2.64(-01) & 9.64(+00) & 5.82(+00) & 2.01(-01) & 1.63(-02) & 7.29(-03) & 1.42(-04) \\
			\\ {$^{68}$Co} & 1 & 8.11(+00) & 2.23(+01) & 4.22(+00) & 2.69(+00) & 2.97(+00) & 2.70(+00) & 1.05(-78) & 3.24(-78) & 1.00(-100) \\
			& 2 & 1.49(+01) & 3.91(+01) & 4.59(+00) & 4.99(+00) & 6.43(+00) & 2.96(+00) & 2.82(-40) & 5.22(-40) & 1.00(-100) \\
			&3 & 1.95(+01) & 5.02(+01) & 4.73(+00) & 6.65(+00) & 9.12(+00) & 3.07(+00) & 3.07(-27) & 4.71(-27) & 2.60(-22)  \\
			&5 & 2.37(+01) & 6.08(+01) & 4.81(+00) & 8.51(+00) & 1.31(+01) & 3.16(+00) & 1.33(-16) & 1.88(-16) & 5.81(-14) \\
			& 10 & 2.90(+01) & 1.35(+02) & 6.84(+00) & 1.29(+01) & 5.28(+01) & 4.89(+00) & 3.00(-08) & 1.13(-07) & 3.38(-07) \\
			& 30 & 1.26(+02) & 2.56(+03) & 8.41(+01) & 1.06(+02) & 2.18(+03) & 7.73(+01) & 1.45(-01) & 2.96(+00) & 3.13(-01) \\
			\\	\hline
	\end{tabular}}
\end{table*} 

\end{document}